\documentclass[10pt]{article}
\usepackage{graphicx}
\usepackage[english]{babel}
\usepackage{amsfonts,latexsym}
\usepackage{amsmath}
\usepackage{epstopdf} 
\usepackage{latexsym}
\usepackage[latin1]{inputenc}

\def\be{\begin{equation}}
\def\ee{\end{equation}}
\def\ba{\begin{eqnarray}}
\def\ea{\end{eqnarray}}
\def\parderiv#1#2{\frac{\partial #1}{\partial #2}}
\def\Vvec{{\bf V}}
\def\rvec{{\bf r}}

\begin{document}

\title{Relativistic Theory of the Falling Retroreflector Gravimeter$^*$}

\author{Neil Ashby$^{1,2}$ \\ $^1$National Institute of Standards and Technology\\
$^2$University of Colorado
Boulder, CO USA}
\maketitle

\begin{abstract}

We develop a relativistic treatment of interference between light reflected from a falling cube retroreflector in the vertical arm of an interferometer, and light in a reference beam in the horizontal arm.  Coordinates that are nearly Minkowskian, attached to the falling cube, are used to describe the propagation of light within the cube.  Relativistic effects such as the dependence of the coordinate speed of light on gravitational potential, propagation of light along null geodesics,  relativity of simultaneity, and Lorentz contraction of the moving cube, are accounted for. The calculation is carried to first order in the gradient of the acceleration of gravity.   Analysis of data from a falling cube gravimeter shows that the propagation time of light within the cube itself causes a significant reduction in the value of the acceleration of gravity obtained from measurements, compared to assuming reflection occurs at the face.  An expression for the correction to $g$ is derived and found to agree with experiment. Depending on the instrument, the correction can be several microgals, comparable to commonly applied corrections such as those due to polar motion and earth tides.  The  controversial ``speed of light" correction is discussed.

\end{abstract}


\vspace{2pc}
\noindent{\it Keywords}: gravimeters, relativity, acceleration, retroreflectors

\renewcommand{\thefootnote}{\fnsymbol{footnote}}
\setcounter{footnote}{1}
\footnotetext{Work of the U.S. government, not subject to copyright.}
\setcounter{footnote}{0}

\section{Introduction}

  In a typical falling cube gravimeter, whose purpose is to measure the acceleration of gravity $g$ with precision, a reference beam passes through a beam splitter where part of the beam is sent vertically to a falling corner cube reflector in an evacuated chamber.  The cube is accelerated by gravity so the reflected beam suffers first- and second-order Doppler shifts as well as other effects; upon mixing with the reference beam back at the beamsplitter a beat frequency is generated having a rapidly increasing frequency chirp.  In the gravimeter analyzed in this paper, a time stamp is recorded repeatedly after some fixed increment in the number of zero crossings of the beat signal.  The time series depends on the strength $g$ of the acceleration of gravity and on its gradient, on the structure of the retroreflector, and the position and velocity of the cube at the initial instant of release; the initial position and velocity and $g$ are then extracted from the data. Typically a great many drops are averaged to obtain the value of $g$; fractional uncertainties of the order of $10^{-9} {\rm cm/s}^2$  are currently obtained after applying several corrections that are of  microgal order; ($1 {\rm\ Gal}=1\ {\rm cm/s}^2)$. 
  
  A freely falling, locally inertial system  can provide an application of the Principle of Equivalence, in which the linear contribution to the gravitational potential is transformed away and the residual gravitational field consists only of tidal terms, or gravity gradients.  The cube is an extended body with more mass near the flat face, so the net gravitational force on the body acts at a point above the face.  For an ideal cube, the center of mass is at a point $d=D/4$ from the face, where $D$ is the cube depth from face to corner.  Choosing the retroreflector's center of mass as the origin of local coordinates simplifies the metric tensor in the freely falling frame, but complicates the analysis of light propagation within the cube.
  
 	Analysis of the light within the cube can be described as though a reference frame fixed in the falling cube were inertial; this entails a small time delay, during which the cube continues to accelerate in the laboratory.   A wave front with a particular phase that enters the cube then leaves the cube at a lesser height so that there are fewer observed fringes in the interference pattern, than if reflection occurred at the face.  The extra path can amount to several thousand wavelengths.  A smaller measured value of $g$ results when this is accounted for.  
 	
 	We construct a reference frame fixed in the cube by the use of normal Fermi coordinates, which are very close to the Minkowski coordinates of special relativity.  We derive a simple expression for the correction to the measured value of $g$ that accounts for the dimensions and refractive index of the cube; the correction can amount to several microgals, comparable to many other commonly applied corrections such as those due to polar motion, and earth tides.

The purpose of the present article is to provide a relativistic theory of a falling cube gravimeter, which fully respects relativistic principles such as the equivalence principle, light propagation along null geodesics, relativity of simultaneity and Lorentz contraction between moving frames.  Tan {\it et al.}\cite{tan16} have used normal Fermi coordinates to treat relativistic effects in an absolute gravimeter; they have studied earth rotation effects, and motion of the falling mass perpendicular to the laser beam; neither of these phenomena are investigated here.  They have not, however, treated the optical path of laser light in the cube, which is the most important topic discussed here. In this paper we account for the phase change of the reflected beam due to its propagation in the glass.  We process data from 5000 drops in a gravimeter of this type, both with and without the assumption that reflection occurs at the face, and find a difference of several  microgals in the value of $g$.   An additional issue treated here in full is dependence of the coordinate speed of light on gravitational potential and the time delays of the test beam upward from the beamsplitter, through the glass, and down to the beamsplitter.
 
In Sect. 2 we discuss the action of a gravity field gradient on a cube of pyramidal shape.  Sect. 3 applies the result to the construction of a cube-fixed locally inertial frame; this work is supported by detailed derivations in the Appendix.  In Sect. 4 the coordinate speed of light is shown to depend on the gravitational potential, and the phase of the upward-propagating test beam reflected from the beam splitter is calculated.  Sect. 5 completes the derivation of the test beam phase and its interference with the reference beam.  Results of data analysis of 5000 drops are discussed in Sect. 7.  The ``speed of light" correction is briefly discussed in Sect. 8.

\section{Free Fall of an Extended Body}

	For a freely-falling point mass, the gravitational potential gradient at the position of the mass is equal to (except for a negative sign) the mass's acceleration.  Construction of local normal Fermi coordinates with origin at the mass's position, results in the elimination of linear terms in the gravitational potential, so that in  the local coordinate system the mass is not accelerated; only quadratic (tidal) terms in the gravitational potential remain.  This is a consequence of the Equivalence Principle: the effective gravitational field, $-{\bf a}$, induced by the acceleration, cancels the real gravitational field so that the mass is ``weightless" in the local coordinate system.  For an extended body, however, the net gravitational force may not act at the center of mass.  Then if the center of mass were chosen as the origin of body-fixed coordinates, one could not expect linear terms in the Taylor expansion of the gravitational potential to be cancelled out as a result of the transformation to the local frame.  
	
	The falling corner cube is an example of a special situation in which the net gravitational force, including the gravity gradient, acts at the center of mass.  The linear mass density of a corner cube of depth $D$, along the symmetry axis, increases quadratically with distance from the corner. We use capital letters to denote coordinates in the laboratory frame.  We assume that the gravitational potential is 
\be\label{potentialinlab}
\Phi(Z)=g Z-\gamma Z^2/2,
\ee	
with the gravity gradient parameter $\gamma \approx 3\times 10^{-6} {\rm s}^{-2}$ near earth's surface.   The gradient parameter $\gamma$ represents the rate of decrease of $g$, per meter of vertical distance.  Let $dM$ be the mass in a slice of retroreflector material between $Z$ and $Z+dZ$.   The gravitational force on $dM$ will be $(-g+\gamma Z)dM$.  Let the cube be placed at rest with its face at $Z$ and its corner at $Z+D$.  The center of mass $\overline{Z}  $ is defined by
\be
M \overline{Z}=\int Z dM,
\ee
so the net force per unit mass on the retroreflector is 
\be 
g-\gamma \overline{Z}=-\frac{\partial \Phi}{\partial Z}\bigg|_{Z=\overline{Z}}\,.
\ee
This suggests that the center of mass, at distance $d$ above the face, should be chosen as the origin of locally inertial coordinates.   For a perfect cube with face ground normally to the $(1,1,1)$ axis, $d=D/4$.
	
\section{The Falling Cube}
	
	Fig. 1 illustrates the operation of a gravimeter that is analyzed in this paper. 
\begin{figure}\label{fig1}
\centering
\includegraphics[width=4.6 truein]{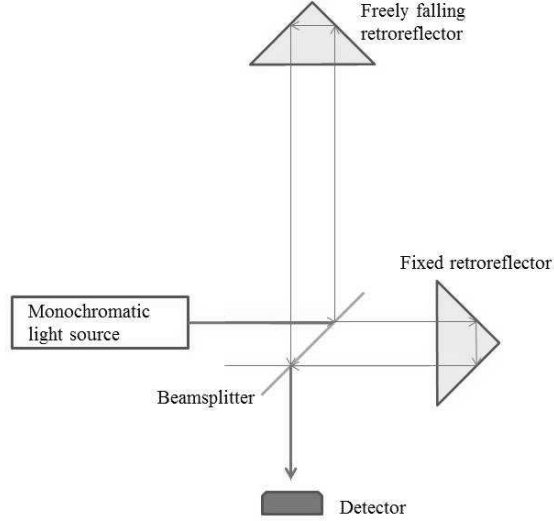}
\caption{Simplified diagram of a freely falling cube gravimeter.}
\end{figure}
	An ideal cube is diagrammed schematically in Fig. 2, showing an incoming ray, followed by three total internal reflections, and an exiting ray.
\begin{figure}\label{fig2}
\centering
\includegraphics[width=2.0 truein]{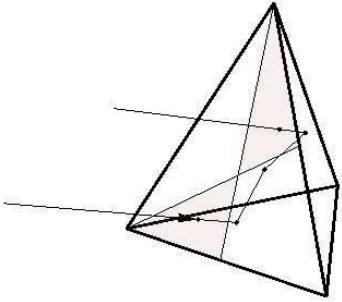}
\caption{An ideal corner cube showing an entering ray, three reflections, and an exiting ray.  Most such retroreflectors will have the points near the face ground away, so the center of mass is closer to the corner than is the case for the cube shown here.}
\end{figure}
	We take as the reference point and origin of laboratory coordinates the point where the beams recombine.  We assume the beamsplitter is perfectly aligned so that the optical path differences in vacuo in the two arms, from the point where the beam is split, to the recombination point, are compensated and do not have to be considered explicitly.  The cube is assumed to be perfectly aligned and to fall without rotation.  In the laboratory the gravitational potential is given by Eq. (\ref{potentialinlab}).  We use lower case letters to denote coordinates fixed in the frame of the falling cube, and take the center of mass of the falling cube to be the origin of coordinates $z=0$ in the falling frame.  Both $Z$ and $z$ are measured positive upwards.
The equation of motion of the cube, measured at the center of mass, is
\be\label{eqofmotion}
\ddot Z = -g+\gamma Z.
\ee
Throughout this paper we neglect all terms that are of order $\gamma^2$ or higher. The solution corresponding to release from position $Z_0$ of the center of mass, with initial velocity $V_0$ at $T=0$ is
\be\label{Zcm}
Z(T)=Z_0+V_0 T-\frac{g T^2}{2}+\gamma\bigg(\frac{Z_0T^2}{2}+\frac{V_0 T^3}{6}-\frac{g T^4}{24}\bigg) \,.
\ee
The velocity of the center of mass is
\be\label{Vcm}
V(T)=V_0-g T+\gamma\bigg(Z_0 T+\frac{V_0 T^2}{2}-\frac{g T^3}{6}\bigg)\,.
\ee
Additional terms arising from solution of relativistic equations of motion give contributions of higher order in  $c^{-1}$ and can be neglected.

To fully describe the physics within the falling cube we need to introduce a transformation relating coordinates and time between the laboratory frame  $(T,Z)$, and the accelerating falling frame $(t,z)$.  We begin by computing the proper time  $\tau$ of a reference clock at the center of mass, elapsed from the drop instant $T=0$.  

Let the metric tensor in the laboratory frame be:
\ba\label{metricinlab}
G_{00}=-\big(1+\frac{2\Phi(Z)}{c^2}\big);\\
G_{ZZ}=\big(1-\frac{2\Phi(Z)}{c^2}\big)\,,
\ea
where $c$ is the speed of light at $Z=0$ determined with the aid of an atomic reference clock.   The potential is assumed to be static in the laboratory frame during the time required for one drop.  The fundamental scalar invariant is 
\be
ds^2=( c d\tau)^2=-G_{\mu\nu}dX^{\mu}dX^{\nu}\,.
\ee
We have adopted the metric signature ${-1,1,1,1}$ so $dt^2>0$; repeated Greek indices are summed from 0 to 3, except that only two components of the metric tensor need to be considered.  Transverse motion is neglected in this paper; we do not consider Coriolis forces or rotation of the retroreflector.

The proper time of a falling clock at the origin of cube-fixed coordinates is then
\ba\label{propertime}
\tau=\frac{1}{c}\int_0^T\sqrt{-(G_{00} (c\, dT)^2+G_{ZZ} dZ^2)}dT\nonumber \hbox to 1in{}\\
\approx \frac{1}{c}\int_0^T\bigg(1+\frac{\Phi(Z(T))}{c^2}-\frac{V(T)^2}{2c^2}\bigg) c\,dT \hbox to 1in{} \nonumber\\
=T\bigg(1+\frac{1}{c^2}\big(Z_0 g -\frac{V_0^2}{2}\big)\bigg)+\frac{V_0 g }{c^2}T^2-\frac{g^2 T^3}{3c^2}\notag\\
+\frac{\gamma}{c^2}\bigg(-\frac{Z_0^2 T}{2}-Z_0V_0T^2\hbox to 1in{}\notag\\
+\big(-\frac{V_0^2}{3}+\frac{2 g Z_0}{3}\big)T^3
+\frac{gV_0T^4}{3}-\frac{g^2 T^5}{15}  \bigg)\,,
\ea
where here and throughout the paper we keep the leading relativistic correction terms of order $c^{-2}$, but keep only the contributions linear in $\gamma$.  

This has an application to atomic fountain clocks, since Eq. (\ref{propertime}) applies to an atom launched upwards.  For atomic fountains, we neglect $\gamma$.  Suppose an atom is projected upwards from $Z_0=0$ with velocity $V_0$ sufficient to reach height $H$, with $V_0^2=2 g H$.  A total time $2V_0/g$ is required for the atom to fall back down to the starting point.  During this interval the proper time elapsed on the atom from Eq. (\ref{propertime}), is
\be 
\tau=\frac{2 V_0}{g}\big( 1+\frac{V_0^2}{6 c^2}\big)\,.
\ee
The ratio of the relativistic part of this, to the non-relativistic part, is just the fractional frequency shift of the hyperfine splitting of the atom, due to relativistic effects, and is
\be 
\frac{\Delta \tau}{\tau}=\frac{\Delta f}{f}=\frac{V_0^3/(3 g c^2)}{2V_0/g}=\frac{V_0^2}{6 c^2}=\frac{1}{3}\frac{g H}{c^2}\,.
\ee
Including $\gamma$ in this calculation yields only tiny additional corrections.

Eq. (\ref{propertime}) for the proper time can be solved for $T$ in terms of $\tau$ at the center of mass by iteration:
\begin{multline}\label{Toftauatcm}
T=\tau\bigg(1+\frac{V_0^2}{2c^2}-\frac{g Z_0}{c^2}  \bigg)-\frac{g V_0\tau^2}{c^2}+\frac{g^2\tau^3}{3 c^2}\\
+\frac{\gamma}{c^2}\bigg(\frac{Z_0^2\tau}{2}+Z_0V_0\tau^2+\frac{(V_0^2-2 g Z_0)}{3}\tau^3
-\frac{g V_0 \tau^4}{3}+\frac{g^2 \tau^5}{15} \bigg)\,.
\end{multline}
As can be seen, the derivations are straightforward but the expressions are algebraically lengthy.  Therefore the complete calculation of the transformation equations, including the transformation function $Z(t,z)$, is provided in Appendix A and we proceed mostly by quoting the results of those calculations. 

In addition to the terms in Eq. (\ref{Toftauatcm}), there is a correction  $V(T)z/c^2$ arising from the relativity of simultaneity.  Adding this term (see Appendix A)  gives the net time transformation
\begin{multline}\label{Ttran}
T(t,z)=t\bigg(1+\frac{V_0^2}{2c^2}-\frac{g Z_0}{c^2}- \frac{g z}{c^2} \bigg)+\frac{V_0 z}{c^2}-\frac{g V_0 t^2}{c^2}+\frac{g^2 t^3}{3 c^2}\hbox to .3in{}\\
\quad\quad+\frac{\gamma}{c^2}\bigg(\big(\frac{Z_0^2}{2}+Z_0 z \big) 
t+\big(V_0Z_0+\frac{V_0 z}{2}\big) t^2 \hbox to .8truein{}\\
\quad\quad\quad\quad+\big(\frac{V_0^2}{3}-\frac{g z}{6}-\frac{2 g Z_0}{3}\big)t^3-\frac{g V_0 t^4}{3}+\frac{g^2 t^5}{15} \bigg)\,.\hbox to 1.in{}
\end{multline}
The symbol $\tau$ is reserved for the proper time on an ideal clock at the origin of falling coordinates, while $t$ represents the time variable extended to the entire region of interest.  At the center of mass, $z=0$ and $t=\tau$.  

The transformation for coordinate $Z(t,z)$ is derived in the Appendix and is
\begin{multline}\label{ZtranfromAppendix}
Z(t,z)=Z_0+z(1+\frac{V_0^2}{2 c^2}+\frac{g Z_0}{c^2})+\frac{g z^2}{2 c^2}\\
{}+t(V_0 +\frac{V_0^3}{2 c^2}-\frac{g V_0 Z_0}{c^2})+t^2(-\frac{g}{2}-\frac{3 g V_0^2}{2 c^2}+
\frac{g^2 Z_0}{c^2})+\frac{4 g^2V_0 t^3}{3 c^2}-\frac{g^3 t^4}{3 c^2}\\
{}+\gamma \bigg(-\frac{Z_0^2 z}{2 c^2}-\frac{Z_0 z^2}{2 c^2}+t(\frac{Z_0^2 V_0}{2 c^2} -\frac{V_0 z^2}{2 c^2})
+t^2(\frac{Z_0}{2}+\frac{3 V_0^2Z_0}{2 c^2}-\frac{3 g Z_0^2}{2c^2}+\frac{g z^2}{4 c^2})\\
{}+t^3(\frac{V_0}{6}+\frac{7 V_0^3}{12 c^2}-\frac{19 g Z_0 V_0}{6 c^2})
+t^4(-\frac{g}{24}-\frac{5 g V_0^2}{4c^2}+\frac{7 g^2 Z_0}{6 c^2})\\
{}+\frac{11 g^2 V_0 t^5}{15 c^2}-\frac{11 g^3 t^6}{90 c^2}\bigg)\hbox to 2 truein{}.
\end{multline}
It is shown in the Appendix that these transformations eliminate the term in $g$ in gravitational potential in the falling frame, with a small contribution remaining from the gravity gradient; in the falling frame the gradient contribution to the gravitational potential is $-\gamma z^2$ where $z$ is measured upwards from the center of mass. The term $ \gamma z^2 $ is precisely the contribution needed for the cube to remain  unaccelerated in the falling frame.  The inverses of these transformations are given in the Appendix.

\section{Signal phase and propagation speed}

	The output signal of the gravimeter can be analyzed in several ways.  One way is to follow the frequency from source through the beamsplitter, through the cube and back down to the beamsplitter where it is combined with the reference beam.  Another way is to follow the phase of a monochromatic signal through the apparatus.  Another way is to imagine sharp pulses emitted from the source, and to calculate the propagation time to the recombination point.  One useful fact in such analyses is, because the gravitational field is static, coordinate frequency during propagation up or down through the gravitational field is conserved.  Another useful fact is: the wavevector of a monochromatic wave is a null fourvector in vacuum. The approach we choose here is to follow the phase of the test signal from the beamsplitter, through the cube, and back down to the beamsplitter.  This is simpler in some respects since phase is a relativistic invariant.  We shall neglect dispersion and make no distinction between group and phase speeds of signals.  Also, we are justified in treating propagation of signals in the cube neglecting gravitational potentials, provided that analysis is done in the freely falling frame.  
	
	 In the lab, a signal propagating in a vacuum in the upwards direction is null and at each point along its path satisfies:
\be
0=G_{00} c^2 dT^2 +G_{ZZ} dZ^2.
\ee
The coordinate speed of the signal will therefore be:
\be\label{coordinatespeed}
V_s=\frac{dZ}{dT}=\sqrt{-G_{00}/G_{ZZ}} \approx c\big(1+\frac{2\Phi(Z)}{c^2}\big). 
\ee
We choose the reference point to be at the beam splitter and set $\Phi(0)=0$ there;  the speed of light at this point is assumed to be the defined speed, $299792458$ m/s.  The coordinate speed above the reference point will be greater than $c$.
	
	Suppose the angular frequency of the reference signal is $\Omega$, and that the signal from the reference beam that stikes the splitter and is reflected up has phase $\phi(T_0)=-\Omega T_0$ at the origin $Z=0$.  The wavelength of the light will be $\Lambda=2 \pi c/ \Omega$.  The phase propagates upwards with coordinate speed given by Eq. (\ref{coordinatespeed}), and will reach height $Z$ at time $T$ given by
\ba
T=T_0+\int_0^Z\frac{dZ}{V_s}=T_0+\frac{1}{c}\int_0^Z\bigg(1+\frac{2 g Z}{c^2}-\frac{\gamma Z^2}{c^2} \bigg)dZ\nonumber\\
=T_0+\frac{1}{c}\bigg(Z+\frac{g Z^2}{c^2}-\frac{\gamma Z^3}{3c^2}  \bigg)\,.
\ea
The phase of the signal at $(T,Z)$ is therefore
\be
\phi(Z,T)=\phi(0,T_0)=-\Omega\bigg(T-\frac{1}{c}\bigg(Z-\frac{g Z^2}{c^2}+\frac{\gamma Z^3}{3c^2}  \bigg)\bigg)\,.
\ee
The signal four-vector $K_{\mu}$ can be obtained from this phase by differentiation:
\be\label{labfourvector}
K_{\mu}=\parderiv{\phi(Z,T)}{X^{\mu}},
\ee
where $X^0=cT$, $X^3=Z$.  This is a null fourvector satisfying
\be\label{null4vectorinlab}
G^{00}(K_0)^2+G^{ZZ}(K_Z)^2=\frac{(K_0)^2}{G_{00}}+\frac{(K_Z)^2}{G_{ZZ}}=0.
\ee
The wavevector components obtained from Eq. (\ref{labfourvector}) are consistent with Eq. (\ref{null4vectorinlab}) to the order $c^{-2}$ of the present calculation.  In particular, the coordinate frequency $K_0=-\Omega/c$ in the lab is conserved in the static gravitational field.

\section{Phase in the falling cube}

We now follow the phase of the signal through the cube until it exits going in the $-z$ direction.  The phase of the signal impinging on the front face of the falling cube is denoted by $\phi_{in}$ and will be
\be\label{phasein0}
\phi_{in}=\phi(Z(t,-d),T(t,-d))
\ee
where $Z(t,-d))$ is the position of the face at time $T(t,-d)$ given by the transformation equations (\ref{ZtranfromAppendix},\ref{Ttran}).  This substitution naturally gives the phase entering the cube in terms of the time $t$ in the local frame.   Substituting and expanding to order $c^{-2}$, the signal phase at the retroreflector face is
\begin{multline}\label{phaseinincube}
\phi_{in}(t)=\Omega\bigg(\frac{Z_0-d}{c}\bigg)+\Omega t\bigg(-1
+\frac{V_0}{c}
-\frac{dg}{c^2}
-\frac{V_0^2}{2c^2}
+\frac{g Z_0}{c^2}\bigg)
+\Omega t^2\bigg(-\frac{g}{2c}+\frac{g V_0}{c^2}  \bigg)\\
+\frac{\Omega d V_0}{c^2}-\frac{\Omega t^3 g^2}{3c^2}
+\gamma \Omega \Biggl(t(-\frac{ Z_0^2}{2 c^2}+\frac{d Z_0}{c^2})
+\frac{Z_0}{2c}
+t^2\bigg(+\frac{d V_0}{2 c^2}
-\frac{V_0Z_0}{c^2}\bigg)\\
+t^3\bigg(
+\frac{V_0}{6c}
-\frac{d g}{6 c^2}
-\frac{V_0^2}{3c^2}
+\frac{2 g Z_0}{3c^2}\bigg)
+t^4\bigg(-\frac{g}{24c}+\frac{g V_0}{3 c^2}\bigg)-\frac{t^5 g^2}{15 c^2}\Biggr)\,.
\end{multline}

The value of $\phi$ in Eq. (\ref{phaseinincube}) is labeled with a subscript ``in" since it corresponds to the phase that strikes the cube face.  The phase is a relativistic invariant, so this is the phase entering the retroreflector in the local freely falling frame.  In this frame (but in vacuum outside the glass) at this point, the coordinate speed of light differs slightly from $c$ because $z=-d$ and the metric tensor still has gravity gradient terms.  However, the difference is negligible, see (\ref{g001}),(\ref{gzz}). We shall neglect dispersion in the cube and assume  that the coordinate phase speed of light in the glass is reduced by a factor $1/n$.  The phase speed is therefore
\begin{equation}\label{eq23}
v_z=\frac{dz}{dt}=\frac{c}{n}\sqrt{-g_{00}/g_{zz}}=\frac{c}{n}\big( 1-\frac{\gamma z^2}{c^2}\big)\,.
\end{equation}
In the falling frame, the local time interval needed to reach height $z$ going upwards is then
\be\label{eq24}
\delta t=\int_{-d}^{z} \frac{dz}{v_z}\approx\frac{n}{c}\int_{-d}^{z} dz \big(1+\frac{\gamma z^2}{c^2} \big)=\frac{n}{c}(z+d+
\frac{\gamma}{3c^2}(z^3+d^3)\big)\,.
\ee
We denote the distance from the face at $z=-d$ to the corner by $D$.  The total local time required for the phase front to propagate to the corner and back to the face is then double the amount calculated just above:
\be\label{deltauplus}
\delta t_+=\frac{2n}{c}\bigg(D+\frac{2\gamma}{c^2}(d^2D-dD^2+\frac{1}{3}D^3)\bigg)\,.
\ee
This time interval is the same for every ray entering the face normally.  The phase $\phi_{out}$ exiting the face is numerically equal to $\phi_{in}$, but the local time at which the phase front exits is at a later local time.  Therefore the replacement
\be\label{phiout}
\phi_{out}=\phi_{in}\vert_{t\rightarrow t-\delta t_+}
\ee
will give an outgoing phase that represents the phase at a later $t$.  In making this replacement, the last terms in $\gamma$ in Eq. (\ref{deltauplus}) do not contribute, to the order of the present calculation, so
\begin{multline}\label{eq27}
\phi_{out}=-\frac{\Omega d}{c}
+\frac{\Omega Z_0}{c}
+\frac{2Dn\Omega}{c}
-\frac{2 D n \Omega V_0}{c^2}
+\frac{\Omega d V_0}{c^2}\\
+t\Omega(-1+\frac{ V_0}{c}-\frac{ V_0^2}{2 c^2}+\frac{2 Dng  }{c^2}+\frac{g Z_0}{c^2}-\frac{ d g}{c^2})\\
+t^2\Omega(-\frac{g}{2c}+\frac{g V_0}{c^2})-\frac{\Omega g^2 t^3}{3 c^2}
+\gamma\Omega\bigg( t(-\frac{Z_0^2}{2 c^2}-\frac{2Dn Z_0}{c^2}+\frac{d Z_0 }{ c^2})\\ +t^2(\frac{Z_0}{2c}+\frac{d V_0}{2c^2}-\frac{V_0Z_0}{c^2}-\frac{D n V_0}{c^2})
+t^3(\frac{V_0}{6c}-\frac{dg}{6c^2}+-\frac{V_0^2}{3 c^2}+\frac{2 g Z_0}{3 c^2}+\frac{D n g}{3 c^2})\\
{}+t^4(-\frac{g}{24 c}+\frac{g V_0}{3 c^2})-t^5\frac{g^2}{15 c^2}\bigg)\,.
\end{multline}
Viewed from the falling frame, the phase front propagates in vacuum downward to negative values of $z$ with a phase velocity 
\be\label{eq28}
v_z=-c(1-\frac{\gamma z^2}{c^2})\,.
\ee
The local time required to propagate to $z<-d$ is
\be\label{eq29}
\delta t_-=-\frac{1}{c}\int_{-d}^z dz(1+\frac{\gamma z^2}{c^2})=-\frac{1}{c}(z+d+\frac{\gamma( z^3+d^3)}{3 c^2})\,.
\ee
The phase field in the region below the cube is then
\be 
\phi(t,z)=\phi_{out}\vert_{t\rightarrow t-\delta t_-}\,.
\ee
Substituting and expanding to order $c^{-2}$ gives
\begin{multline}\label{eq31}
\phi(t,z)=\frac{\Omega Z_0}{c}+\frac{2 D n \Omega}{c}+z(-\frac{\Omega}{c}+\frac{\Omega V_0}{c^2})-\frac{2 D n \Omega V_0}{c^2}\\
+{}t(-\Omega+\frac{\Omega V_0}{c}+\frac{2 D g n\Omega}{c^2}-\frac{\Omega V_0^2}{2 c^2}+\frac{g\Omega Z_0}{c^2}-\frac{g\Omega z}{c^2})
+t^2(-\frac{2 d Dn}{c^2}-\frac{g\Omega}{2c}+\frac{g\Omega V_0}{c^2})\\
-t^3\frac{g^2\Omega}{3 c^2}+\gamma\Omega\big(t(-\frac{2 d Dn}{c^2}-\frac{2 D n Z_0}{c^2}-\frac{ Z_0^2}{2 c^2}++\frac{d z}{c^2}+\frac{ Z_0 z}{c^2}) \\ 
{}+t^2(\frac{d}{2c}-\frac{d V_0}{2c^2}+\frac{ Z_0}{2c}-\frac{D n  V_0}{c^2}-\frac{ V_0 Z_0}{c^2}+\frac{ V_0 z}{2 c^2})\\
{}+t^3(\frac{dg}{2c^2}+\frac{D g n}{3 c^2}+\frac{V_0}{6c}-\frac{V_0^2}{3 c^2}-\frac{g z}{6 c^2}+\frac{2 g Z_0}{3}c^2)\\
+t^4(-\frac{g}{24c}+\frac{g V_0}{3 c^2})-t^5\frac{g^2 }{15 c^2}\big)
-\frac{\gamma \Omega n D^2 t}{2 c^2}\,.
\end{multline}
This phase needs to be evaluated at the reference point $Z=0$ and expressed as a function of the time $T$.  Substituting from the transformation equations Eq. (\ref{tautrana}) and (\ref{zedtran}) and then setting $Z=0$ yields the phase to be combined with the reference phase at the splitter: 
\begin{multline}\label{finaltestphase}
\phi(T,0)=\frac{2( D n-d) \Omega}{c}
+\frac{2 \Omega Z_0}{c}
-\frac{2 (D n-d) \Omega V_0}{c^2}
-\frac{2 \Omega Z_0 V_0}{c^2}\\
{}+T(-\Omega +\frac{2\Omega V_0}{c}-\frac{2\Omega V_0^2}{c^2}+\frac{2\Omega g ( D n-d)}{c^2}+\frac{2 g \Omega Z_0}{c^2})\\
+T^2(-\frac{g \Omega}{c}+\frac{3 g \Omega V_0}{c^2})-\frac{g^2 \Omega T^3}{c^2}\\
{}+\gamma \Omega \bigg( T(-\frac{2 ( D n-d)Z_0}{c^2}-\frac{2 Z_0^2}{c^2})
+T^2(\frac{Z_0}{c}-\frac{(D n-d) V_0}{c^2}-\frac{4V_0Z_0}{c^2})\\
{}+T^3(\frac{V_0}{3c}
+\frac{g( D n-d)}{ 3c^2}-\frac{4 V_0^2}{3 c^2}
+\frac{7g Z_0}{3c^2})
{}+T^4(-\frac{g}{12 c}+\frac{5 g V_0}{4 c^2})-\frac{g^2 T^5}{4 c^2}  \bigg)
\,.
\end{multline}

To the order of this calculation, solving the relativistic equations for the hyperbolic center of mass motion of the cube gives additional terms of order $c^{-2}$ in Eqs. (\ref{Zcm}) and (\ref{Vcm}).  These terms contribute to the transformation equations but because of their high order have no effect on the final phase, Eq. (\ref{finaltestphase}).

The signal is mixed with the reference signal, which has phase $-\Omega T$, and fringes are counted.  The phase difference is obtained from Eq. (\ref{finaltestphase}) by omitting the term $-\Omega T$ in the second line.   It is possible to absorb some of the terms in the cube depth $D$ in the difference by redefining the initial position $Z_0$, however it is not possible to eliminate all such terms and a significant effect remains, which will be discussed in Sect. 6.

The transformations between laboratory coordinates and freely falling coordinates provide alternative descriptions of phenomena, which must agree when results are expressed in terms of observables such as numbers of interference fringe counts.  For example, in the above calculation of the test beam phase at the detector the calculation treated the source at $z=-d$ as at rest and the detector as moving.  Alternatively, one may calculate the test beam phase at the detector in laboratory coordinates considering the source (the cube face) to be moving while the detector is at rest.  In the latter case, however, one must take care to compute the time delay between the source and arrival at the detector accounting for the finite speed of light.  This is analagous to computing the propagation time of a signal sent from a moving transmitter to a receiver at rest.  

Let the face of the cube at transmission time $T_T$ be $Z_{face}(T_T)$ and let the time required for the signal to reach the detector at $Z=0$ at time $T$ be denoted by $\Delta(Z_{face}(T_T))$.  Then the time of transmission is determined by a retarded time:
\be
T_T=T-\Delta(Z_{face}(T_T).
\ee
This can be solved by iteration:
\be 
T_T=T-\Delta(Z_{face}(T-\Delta(Z_{face}(T-\Delta(Z_{face}(T-...)))))).
\ee
Each iteration introduces terms having one more factor of $c$ in the denominator; convergence is rapid.  

We denote the phase out of the cube as a function of $T_T$ in laboratory coordinates as $\Phi_{out}(T_T)$.  This may be obtained by solving the equation $T_T=T(t,-d)$ from the time transformation, for $t$ in terms of $T_T$ and substituting for $t$ in favor of $T_T$ into the scalar, Eq. (\ref{eq27}).  The phase at the detector is then 
\begin{multline}
\hbox to .3 truein{}\Phi(T)=\Phi_{out}(T_T)\\
=\Phi_{out}(T-\Delta(Z_{face}(T-\Delta(Z_{face}(T-\Delta(Z_{face}(T-...))))))).
\end{multline}
The result agrees with the test beam phase at the splitter, Eq. (\ref{finaltestphase}).

{\it Origin of Time.}  The choice of zero of time is arbitrary; in the present work we have chosen to make $T=0$ at the instant the cube is dropped, introducing initial position $Z_0$ and velocity $V_0$ to account for imperfections in release.  Suppose instead the cube were dropped at a different instant $T_0$.  The center of mass position and velocity at this instant will be: 
\begin{multline}
Z(T_0)=Z_0+V_0 T_0-\frac{1}{2}g T_0^2+\gamma\big(\frac{Z_0 T_0^2}{2}+\frac{V_0 T_0^3}{6}-\frac{g T_0^4}{24} \big);\\
V(T_0)=V_0-g T_0+\gamma\big(Z_0T_0 +\frac{V_0 T_0^2}{2}-\frac{g T_0^3}{6}  \big)\,.
\end{multline}
Solving for the original position and velocity, keeping linear terms in $\gamma$, gives
\begin{multline}\label{posvelsub}
Z_0=Z(T_0)-V(T_0) T_0-\frac{g T_0^2}{2}+\gamma\big(\frac{Z(T_0) T_0^2}{2}-\frac{V(T_0) T_0^3}{6}-\frac{g T_0^4}{24}  \big)\,;\\
V_0=V(T_0)+gT_0 +\gamma \big( -Z(T_0)T_0+\frac{V(T_0) T_0^2}{2}+\frac{g T_0^3}{6} \big)\,.
\end{multline}
If one substitutes the replacements indicated in Eqs. (\ref{posvelsub}) into Eq. (\ref{finaltestphase})  {\it or} one lets $T\rightarrow(T-T_0)$, and makes the replacements $Z_0\rightarrow Z(T_0),\ V_0 \rightarrow V(T_0)$, then upon neglecting powers of $\gamma$ higher than 1 it is found that the interference phase difference is form-invariant with respect to this change of time origin.  For example, consider only the terms proportional to $c^{-1}$ in Eq. (\ref{finaltestphase}).  These are
\begin{multline}\label{termsinoneoverc}
\frac{\Omega}{c}\bigg( -2d+2Dn-g T^2+2 Z_0 +2 V_0 T
+\gamma\big(-\frac{1}{12}gT^4+\frac{1}{3}V_0T^3+Z_0T^2\bigg)\,.
\end{multline}
Making the replacements given in Eq. (\ref{posvelsub}) and keeping terms of order $\gamma$, this becomes
\begin{multline}
\frac{\Omega}{c}\bigg(-2d+2Dn-gT^2+2 g T T_0+g T_0^2+2 T V(T_0)-2 T_0 V(T_0)+2Z(T_0)\\
+\frac{1}{12}\gamma\big(g(T-T_0)^4-4(T-T_0)^2V(T_0) +3 Z(T_0)(T-T_0)^2\big)\bigg)
\,,
\end{multline}
which is of the same form as Eq. (\ref{termsinoneoverc}) upon making the replacements $Z_0\rightarrow Z(T_0)$,     $V_0\rightarrow V(T_0),T\rightarrow T-T_0$.   The remaining contributions in Eq. (\ref{finaltestphase}) behave similarly, as do all the terms proportional to $D$.  This is a useful self-consistency check of the theory.

\section{Dispersion and Modulation}

	{\it Dispersion\/}.  As the cube falls, the wavelength of laser radiation within the cube decreases slightly.    This results in a change of phase velocity that is negligibly small, as can be seen from the following argument.    The acceleration $g$ acts for a time $T$ that is only a few tenths of a second, causing the velocity to build up to $g T \approx 5 {\rm\ m/s}^2$.  The first-order Doppler shift of the laser wavelength within the cube is then no more than $\Delta \lambda =- V\lambda/c =-g T\lambda/c \approx- 10^{-14}$ m. Chromatic dispersion for typical corner-cube glass is $dn/d\lambda\approx 5\times 10^4 {\rm\ m}^{-1}$,\cite{schottglass} so $\Delta n\approx 5\times 10^{-10}$.  The change in time delay within the cube due to dispersion is negligible.
	
	{\it Modulation\/}.   Low-frequency modulation may be applied to the laser signal to aid in locking the laser to a frequency reference--e.g., a Rubidium oscillator. This modulation produces sidebands whose strength depends on the amplitude of the modulation.  Every term in the beat frequency is proportional to the original laser frequency of the source, $\Omega$, so both reference beam and retroreflected beam will have sidebands that undergo a chirp proportional to the frequency chirp of the main signal.  We shall write the unmodulated phase of the signal $\Phi(T,0)$ in (\ref{finaltestphase}) in the form
\be
\Phi(T,0)=\Omega F(Z_0,V_0,g,\gamma,T)=\Omega F\,.
\ee	
	Assuming the signal is frequency modulated with relatively low frequency $\omega_m$, and has a small modulation index $\beta$, the modulated reference beam signal at distance $x$ from the origin  can be represented by
\be 
E_{ref}=E_0e^{(i\Omega x/c-i\Omega T)}\big( 1-\frac{\beta}{2}e^{(-i x \omega_m+i \omega_m T)}
																			+\frac{\beta}{2}e^{(i x \omega_m-i \omega_m T)}\,.
\ee
The electric field of the retroreflected beam may have a slightly attenuated amplitude, but will have sidebands such that
\be
E_{test}=E_1e^{i\Omega F}\big(1-\frac{\beta}{2}e^{i\omega_m F}+\frac{\beta}{2}e^{-i\omega_m F}	\big)\,.
\ee
These signals are superimposed and sent into a photodiode.  The measurements consist of time stamps of zero-crossings of the time varying quantity
\be 
E_{test}E_{ref}^*+c.c\,.
\ee
The sideband frequencies also suffer from the frequency chirp that occurs as the cube falls and can be described by the phase function Eq. (\ref{finaltestphase}) with an appropriate frequency.  However, unless the frequency modulation index is large we find there is almost no effect on the interference fringe counts.

\section{Data Analysis}

	Three data sets produced by a falling cube gravimeter were analyzed.  These were characterized by differing numbers of drops, and differing numbers of zero-crossings of the interference signal between time stamps.  These details are summarized in Table 1.
	
	The data were processed using the result in Eq. (\ref{finaltestphase}), with the term $-\Omega T$ in the second line omitted. An estimated value of gravity gradient $\gamma=3.0724615 \times 10^{-6} {\rm\ s}^{-2}$ was used for all drops; the data is not sufficiently robust to determine $\gamma$ itself as the covariance matrix becomes singular if $\gamma$ is included as a variable to be determined by the drop data.  For each drop in each ``project," the data were first processed with $D=0$, as though the reflection from the falling cube occurred at the cube face.  The variables $Z_0, V_0$, and $g$ were determined by the fit with a fixed value of $\gamma$.  Then the data were processed with $D=0.0175$ m corresponding to the distance from cube face to corner of a typical ``1-inch" cube, and $d=D/4$.  (The values of $D$ and $d$ for the particular gravimeter from which this data were taken was not available.)  None of the usual corrections such as those arising from polar motion or tides were applied since such corrections would be the same whether $D$ was or was not included in the calculation. The value of $g$ obtained when $D\ne 0$  was invariably smaller than the value obtained when $D=0.0175$ m.  In fact the difference for all drops in all projects, including short drops obtained by selecting fewer time stamps, was found to be
\be\label{deltagfit}
\delta g = g(D=0.0175 \rm{\ m}) - g (D=0 \rm{\ m})=-6.82854\  \mu {\rm Gal}\,.
\ee
with negligible variation.   The same number arises if all terms of order $c^{-2}$ in Eq. (\ref{finaltestphase}) are neglected.

This is a significant correction, comparable to the magnitudes of many other corrections such as arise from polar motion, or earth tides.  

The fact that the difference $\delta g$ was found to be the same for all drops implies this constant difference should be derivable from Eq. (\ref{finaltestphase}). Suppose that fitting the drop data to Eq. (\ref{finaltestphase}) with $D=0$ results in values $Z_0,V_0,g$ for these three parameters.  Then suppose that fitting the drop data with $D\ne0$ results in the values $Z_0+\delta Z_0,V_0+\delta V_0, g+\delta g$.  All these quantities are constants independent of $T$, so the two phase functions at an arbitrary value of T must match.  Let us denote the phase given in Eq. (\ref{finaltestphase}) by $\phi(D,Z_0,V_0,g,T)$.  Linearizing the phase with respect to the small increments $\delta Z_0,\delta V_0,\delta g$ then upon subtracting the phase without $D$ the difference should be identically zero.  Thus we expect
\be\label{diffduetoD}
\phi(D,Z_0+\delta Z_0,V_0+\delta V_0,g+\delta g,T)-\phi(0,Z_0,V_0,g,T)=0.
\ee
Carrying out the calculation, we find neglecting terms of order $\gamma^2$ but without further approximation, that
\ba\label{deltas}
&\delta Z_0=-D n+d;\nonumber\\
&\delta V_0=0;\\
&\delta g =- \gamma (Dn-d).\nonumber
\ea
It is remarkable that these replacements reduce both non-relativistic terms and terms proportional to $c^{-1}$ in Eq. (\ref{diffduetoD}) to negligible levels.   This value of the correction to $g$ is indeed equal to that found by fitting the data:
\be
\delta g=-6.82854\ \mu {\rm Gal}.
\ee
The changes in $Z_0$ and $V_0$ as given in Eq. (\ref{deltas}) are similarly verified by fitting.  Although Eq. (\ref{diffduetoD}) is a fourth-order polynomial involving five coefficients, the adjustment of only three quantities as in Eq. (\ref{deltas}) is sufficient because $\delta V_0=0$ eliminates all odd terms in $T$.  The residual after applying the remaining two adjustments in Eq. (\ref{deltas}) are proportional to $\gamma^2 $ and are negligible.
\begin{table}
\begin{center}	
\begin{tabular}{c c c c}
\hline
\vbox to 10pt{} Project \# &  Number of drops & Number of & Number of zero-crossings\\
& & time stamps& between time stamps \\
\hline
\#1 & 200&1200 &1000 \\
\#2 &2400 &1400 &800\\
\#3 & 2400&1000 &1200 \\
\hline
\end{tabular}
\caption{Description of data processed using Eq. (\ref{finaltestphase}) }
\end{center}
\end{table}

\section{``Speed of light" corrections}
One of the more controversial issues surrounding measurements with falling cube gravimeters is the so-called ``speed of light" correction.  Many measurement models of gravimeters of this type attempt to account for propagation delays between the instant the waves are reflected from the cube and the instant they are delivered to the detector, due to the finite value of the speed of light.   A review of such speed of light corrections has been given in \cite{nagornyi11}.  It should be emphasized that the result given in Eq. (\ref{finaltestphase}) {\it does not require any such correction} as light propagation has been fully described in that expression using relativistic principles. 

Suppose one nevertheless attempts to derive a ``speed of light correction" by analogy with Eq. (\ref{diffduetoD}) by first fitting only with the leading $c^{-1}$ terms, then linearizing.  To compare with earlier treatments we take $D=0$ and separate the phase contributions into $c^{-1}$ and $c^{-2}$ contributions:
\be
\phi(0,Z_0,V_0,g)=\frac{1}{c}\phi_1(Z_0,V_0,g)+\frac{1}{c^2}\phi_2(Z_0,V_0,g)\,.
\ee
Let $Z_0,V_0,g$ be determined by data fitting using only $\phi_1$, the non-relativistic part of the phase.  Then one might expect corrections to satisfy
\be 
\frac{1}{c}\phi_1(Z_0+\delta Z_0,V_0+\delta V_0,g+\delta g)-\frac{1}{c^2}\phi_2(Z_0,V_0,g)=0.
\ee
However, this is a fourth-order polynomial in the time having five constant coefficients and cannot be satisfied by adjusting three fitting parameters.  Even if terms in $\gamma/c$ were neglected, there are still four constant coefficients.  Thus it appears to be infeasible to derive speed-of-light corrections in this way.   In case of a three-level schema in which only three fringe shifts are recorded at three times, one can imagine obtaining relativistic corrections by adjusting only three parameters.

In any case, such corrections are unnecessary; the full relativistic phase difference including propagation delays of light through the apparatus is available in Eq. (\ref{finaltestphase})
  
\section{Conclusions}

	The large value of the correction in Eq. (\ref{deltas}), obtained from data analysis as well as from theory, supports the contention that in falling cube gravimeters, the time delay entailed by penetration of the light into the cube cannot be ignored.  It also points up the need for a better value of the gravity gradient--not just a ``standard" value--at the position of the apparatus, as well as a precise value for the refractive index of the glass and an accurate position for the center of mass.  No additional ``speed of light" correction is needed in this picture, as relativistic corrections including Lorentz contraction of the falling cube, dependence of the coordinate speed of light on gravitational potential, relativity of simultaneity, propagation along null geodesics, and first- and second-order Doppler shifts have been accounted for.   The correction for time delay within the falling cube has been shown to be several microgals and to depend on properties of the cube; this correction can simply be added to other relevant corrections that are commonly applied.
	
	{\it\ Acknowledgements\/}.  We thank Tim Niebauer and Derek van Westrum for discussions and many suggestions.  We are especially indebted to Tim for providing data for 5000 drops from a falling-cube gravimeter.
	
\section{APPENDIX I.  Coordinate Transformations}

The prescription for constructing accelerated cube-fixed normal Fermi coordinates that cover the falling cube, extending to a space-time patch containing the cube and beamsplitter, is developed in \cite{bertotti86}.   Normal Fermi coordinates\cite{fermi22}\cite{manasse63}, similar to Newtonian cartesian coordinates falling with the cube, are constructed by parallel propagation of a tetrad of orthonormal four-vectors along the trajectory of the freely falling test object.  In the present case the trajectory is that of the center of mass of the falling cube.  The $0^{th}$ member of the tetrad is just the four-velocity, tangent to the trajectory. The time in the falling frame is the proper time on an ideal atomic clock carried along at the center of mass. The other members of the tetrad are obtained by solving equations for spacelike geodesics that intersect the trajectory orthogonally.  In the present case the only additional coordinate of interest is the $Z$-coordinate so only two members of the tetrad are relevant.  The construction yields the following expression for the time transformation (Eq. A.12 in \cite{bertotti86}):
\be
T(t,z)=\int_0^{t}\frac{dt}{\sqrt{-G_{\mu\nu}dX^{\mu}dX^{\nu}}}+\frac{\Vvec \cdot \rvec}{c^2}\,.
\ee
The integral has already been calculated in Eq. (\ref{Toftauatcm}).  The term representing relativity of simultaneity is
\ba\label{simulcorrection}
\frac{V(T)z}{c^2}=\frac{1}{c^2}\big(V_0z-g z T+\gamma z (Z_0 T+\frac{V_0 T^2}{2}-\frac{g T^3}{6}\big)\nonumber\\
\hbox to .5in{}=\frac{1}{c^2}\big(V_0 z-g t z+\gamma(Z_0 z t +\frac{V_0 z t^2}{2}-\frac{g t^3 z}{6} \big)\,.
\ea
where to the order of the calculation $T$ can be replaced by $t$ in Eq. (\ref{simulcorrection}).  Combining Eqs. (\ref{Toftauatcm}) and (\ref{simulcorrection}) gives the transformation from time in the lab to the falling coordinates in the cube, Eq. (\ref{Ttran}).

We also need the transformation of vertical coordinates from the lab coordinate $Z$ to the cube-fixed coordinate $z$, which is given by (Eq. A.10 in \cite{bertotti86}):
\begin{multline}\label{Ztran0}
Z=Z(t,z)=\bigg(Z_{cm}(T)+z \big(1-\frac{\Phi(Z(T))}{c^2}-\frac{A(T)z}{c^2}\big)\\
\hbox to 2 truein{}+\frac{V(T)^2z}{2c^2}+\frac{z^2 A(T)}{2 c^2}\bigg)\bigg|_{T=T(t,0)}\,.
\end{multline}
Here quantities such as $Z(T)$ and $\Phi(T)$ are evaluated at the cube's center of mass, which is the origin of locally inertial, freely falling coordinates. Then T is replaced by its value at the center of mass given by Eq. (\ref{Ttran}). The acceleration $A(T)$ of the cube is obtained by differentiating Eq. (\ref{Vcm}).  The first term in Eq. (\ref{Ztran0}) is the accelerating value of the $Z-$coordinate at the center of mass.  The coefficients of $z$ include a change of scale arising from the gravitational potential external to the cube, a Lorentz contraction term, and additional acceleration contributions.  These contributions arise during construction of Normal Fermi coordinates, while solving for spacelike geodesics that intersect the trajectory orthogonally.\cite{bertotti86} In General Relativity, arbitrary coordinate transformations are allowed; of course then the physics in the resulting coordinate system must be interpreted in terms of physical principles.  

After expanding to order $c^{-2}$, the transformation for the $Z-$coordinate is
\begin{multline}\label{Ztran}
Z(t,z)=Z_0+z(1+\frac{V_0^2}{2 c^2}+\frac{g Z_0}{c^2})+\frac{g z^2}{2 c^2}\\
{}+t(V_0 +\frac{V_0^3}{2 c^2}-\frac{g V_0 Z_0}{c^2})+t^2(-\frac{g}{2}-\frac{3 g V_0^2}{2 c^2}+
\frac{g^2 Z_0}{c^2})+\frac{4 g^2V_0 t^3}{3 c^2}-\frac{g^3 t^4}{3 c^2}\\
{}+\gamma \bigg(-\frac{Z_0^2 z}{2 c^2}-\frac{Z_0 z^2}{2 c^2}+t(\frac{Z_0^2 V_0}{2 c^2} -\frac{V_0 z^2}{2 c^2})
+t^2(\frac{Z_0}{2}+\frac{3 V_0^2Z_0}{2 c^2}-\frac{3 g Z_0^2}{2c^2}+\frac{g z^2}{4 c^2})\\
{}+t^3(\frac{V_0}{6}+\frac{7 V_0^3}{12 c^2}-\frac{19 g Z_0 V_0}{6 c^2})
+t^4(-\frac{g}{24}-\frac{5 g V_0^2}{4c^2}+\frac{7 g^2 Z_0}{6 c^2})\\
{}+\frac{11 g^2 V_0 t^5}{15 c^2}-\frac{11 g^3 t^6}{90 c^2}\bigg)\hbox to 2 truein{}\,.
\end{multline}
The relativistic correction terms are easily identified by the factors of $c$ in the denominators.  The inverses of these transformations, obtained by an iterative process, are
\begin{multline}\label{tautrana}
t(T,Z)=\frac{V_0 Z_0}{c^2}-\frac{V_0 Z}{c^2}+T(1+\frac{V_0^2}{2 c^2}+\frac{g Z}{c^2})
-\frac{g V_0 T^2}{2 c^2}+\frac{g^2 T^3}{6 c^2}\\
+\gamma\bigg( T(\frac{Z_0^2}{2 c^2}-\frac{Z_0 Z}{c^2})+T^2(\frac{V_0 Z_0}{c^2}-\frac{V_0 Z}{2 c^2}) \\
+T^3(-\frac{g Z_0}{2 c^2}+\frac{V_0^2}{3 c^2}+\frac{g Z}{6 c^2})
-\frac{7 g T^4 V_0}{24 c^2}
+\frac{7g^2 T^5}{120 c^2}  \bigg)\,;
\end{multline}
\begin{multline}\label{zedtran}
z(T,Z)=-Z_0-\frac{V_0^2 Z_0}{2 c^2}+\frac{g Z_0^2}{2 c^2}+(1+\frac{V_0^2}{2 c^2})Z-\frac{g Z^2}{2 c^2}
-\frac{3g^2T^3 V_0}{2 c^2}+\frac{3g^3 T^4}{8 c^2}\\
{}+T(-V_0-\frac{V_0^3}{2 c^2}+\frac{2 g V_0Z_0}{c^2}-\frac{g V_0 Z}{c^2})
+T^2(\frac{g}{2}+\frac{7 g V_0^2}{4 c^2}-\frac{g^2 Z_0}{c^2}+\frac{g^2 Z}{2 c^2})\\
{}+\gamma\bigg(-\frac{Z_0^2 Z}{2 c^2}+\frac{Z_0Z^2}{2 c^2}
 +T(-\frac{V_0 Z_0^2}{c^2}+\frac{V_0 Z^2}{2 c^2})\\
\hbox to.8in{}+T^2(-\frac{Z_0}{2}-\frac{7 V_0^2 Z_0}{4 c^2}+\frac{3 g Z_0^2}{2 c^2}-\frac{g Z Z_0}{2 c^2}-\frac{g Z^2}{4 c^2})\\
\hbox to .7in{}+T^3(-\frac{V_0}{6}+\frac{10 g V_0 Z_0}{3 c^2}-\frac{7 V_0^3}{12 c^2}-\frac{g V_0 Z}{6 c^2})\\
\hbox to .8in{}+T^4(\frac{g}{24}+\frac{61gV_0^2}{48 c^2}-\frac{29 g^2 Z_0}{24 c^2}+\frac{g^2 Z}{24 c^2})\\
{}-\frac{3 g^2 V_0 T^5}{4 c^2}+\frac{g^3 T^6}{8 c^2} \bigg)\,.\hbox to .3in{}
\end{multline}

These transformations can be used to investigate transformations of laser light wavevectors from the laboratory frame to the cube frame.  Partial derivatives such as
\be
\parderiv{X^i}{x^i}\,,
\ee
where $X^i=\{c T,Z\}$ and $x^i=\{ct,z\}$ can be easily evaluated from the polynomials given in Eqs. (\ref{Ttran}) and (\ref{Ztran}).  For example, the time-time-component of the metric tensor in the falling frame will be
\be\label{g00}
g_{00}=\bigg(\parderiv{cT}{ct}\bigg)^2G_{00}+\bigg(\parderiv{Z}{ct}\bigg)^2 G_{ZZ}\,.
\ee
Carrying out the partial differentiations and evaluating Eq. (\ref{g00}), expanding to order $c^{-2}$,  keeping linear contributions in $\gamma$, and quadratic contributions in $z$, we obtain
\be\label{g001}
g_{00}=-(1-\gamma\frac{z^2}{c^2})\,.
\ee
A similar calculation gives:
\be\label{gzz}
g_{zz}= \bigg(\parderiv{cT}{z}\bigg)^2G_{00}+\bigg(\parderiv{Z}{z}\bigg)^2 G_{ZZ}=1+\gamma\frac{z^2}{c^2}\,.
\ee
There are no linear terms in $g_{00}$ in the local $z-$ coordinate because the origin was chosen to be at the point where the net force acts.  In the cube-fixed frame the center of mass is at
\be
z_{cm}=\frac{1}{M}\int_{-d}^{D-d}z\, dm=0,
\ee
 while the force per unit mass at $z$ is proportional to $-\gamma z$.  The total force is then proportional to
\be 
-\gamma \int_{-d}^{D-d} z\, dm=0,
\ee
so the retroreflector is unaccelerated in the local freely falling frame.   If one should choose the flat face of the falling object as origin of the local frame, the local metric tensor would have linear terms in $z$.

Such features express the equivalence principle: in a freely falling reference frame, the linear term in the gravitational potential is cancelled by terms arising from the acceleration.  This is primarily an algebraic  consequence of the simultaneity term Eq.(\ref{simulcorrection}) that was added into the time transformation Eq. (\ref{Ttran}).   The choice of origin we have made is at the center of mass, because the description of acceleration is simpler; however the integrations involved in analysis of phase advance are more complicated.

Not all of the relativistic correction terms are important; many of the terms multiplying $\gamma$ are very small. Within the falling cube, $z$ is usually no more than a few centimeters; with a nominal value of the gravity gradient at Earth's surface $\gamma \approx 3 \times 10^{-6}$, then $ \gamma z^2/(2c^2) \approx 10^{-26}$ and could be neglected. Nevertheless these terms are included since they may be important in other applications.   Although one may have questions about the transformations quoted in Eqs. (\ref{Ttran}) and (\ref{Ztran}), the result clearly describes the physics in a freely-falling, locally inertial system with a static gravity gradient, with coordinates that are locally nearly Minkowskian.  The linear term in the potential involving $g$ has been transformed away. 

\section{APPENDIX II.  Non-relativistic derivation\\ of corrections.}

The corrections given in Eq. (\ref{deltas}) do not involve the constant $c$ explicitly, and can be derived by taking the non-relavitistic limit of the equations presented above. It is instructive to summarize the argument. For example, the difference between the time coordinates in the lab and in the falling frame can be neglected.  From the transformation equations, the position of the cube face is then approximately
\be
Z_{face}(T)=Z_{cm}(T)-d.
\ee 
The phase of the signal penetrating the cube is then
\be 
\phi_{in}(T)=-\Omega\big(T-\frac{Z_{face}(T)}{c}\big)\,.
\ee
The time delay during signal propagation within the cube is $2 D n/c$ so the phase of the signal leaving the cube is
\begin{multline} 
\phi_{out}(T)=\phi_{in}(T-2 D n/c)\\
=-\Omega\big( T-\frac{2 D n}{c}-\frac{Z_{face}(T-2 D n/c)}{c}\big)\\
\approx -\Omega \big(T-\frac{2 D n}{c}-\frac{Z_{face}(T)}{c} \big)\,,
\end{multline}
where in the last term the higher-order relativistic correction can be neglected.
The time required for the signal to reach the origin $Z=0$, to leading order in $c^{-1}$, is $Z_{face}(T)/c$, so the phase of the signal at the detector is
\begin{multline}
-\Omega\big(T-\frac{Z_{face}(T)}{c}-\frac{2D n}{c}-\frac{Z_{face}(T-Z_{face}(T)/c}{c}\big)\\
=-\Omega \big(T-\frac{2 D n}{c}-\frac{2Z_{face}(T)}{c}  \big)\,.
\end{multline}
Replacing $\Omega$ by $2\pi c/\lambda$, the ``nonrelativistic" contributions to the phase is
\be\label{nrphasefinal}
-\frac{2 \pi c}{\lambda}\big(T-\frac{2Dn}{c}-\frac{2Z_{face}(T)}{c}\big)\,.  
\ee
The first term is removed by interference with the reference beam so the net nonrelativistic phase difference is
\be
\frac{2 D n}{\lambda}+\frac{2 Z_{face}(T)}{\lambda}\,.
\ee
Use of this expression in Eq. (\ref{diffduetoD})
yields the corrections quoted in Eq. (\ref{deltas}).

\end{document}